\documentstyle[doublespace,prl,epsf,aps,floats]{revtex}  
\def\met{\mbox{${\hbox{$E$\kern-0.6em\lower-.1ex\hbox{/}}}_T$}} 
\def\mex{\mbox{${\hbox{$E$\kern-0.6em\lower-.1ex\hbox{/}}}_x$}} 
\def\mey{\mbox{${\hbox{$E$\kern-0.6em\lower-.1ex\hbox{/}}}_y$}} 
\def\mexy{\mbox{${\hbox{$E$\kern-0.6em\lower-.1ex\hbox{/}}}_{x,y}$}} 
 
\def\D0{D\O}                            
\begin{document}
\lefthyphenmin=2
\righthyphenmin=3

%
%
\title{
Search for Doubly-Charged Higgs Bosons Decaying to Dileptons in 
$\boldmath{ p\overline{p}}$ Collisions at $\boldmath{ \sqrt{s}}$=1.96 TeV\\
}

\maketitle

\font\eightit=cmti8
\def\r#1{\ignorespaces $^{#1}$}
\hfilneg
\begin{sloppypar}
\noindent
D.~Acosta,\r {16} T.~Affolder,\r 9 T.~Akimoto,\r {54}
M.G.~Albrow,\r {15} D.~Ambrose,\r {43} S.~Amerio,\r {42}  
D.~Amidei,\r {33} A.~Anastassov,\r {50} K.~Anikeev,\r {31} A.~Annovi,\r {44}
% ...
\end{sloppypar}
\vskip .026in
\begin{center}
(CDF Collaboration)
\end{center}

\vskip .026in
\begin{center}
\r 1  {\eightit Institute of Physics, Academia Sinica, Taipei, Taiwan 11529, 
Republic of China} \\
\r 2  {\eightit Argonne National Laboratory, Argonne, Illinois 60439} \\
\r 3  {\eightit Institut de Fisica d'Altes Energies, Universitat Autonoma
de Barcelona, E-08193, Bellaterra (Barcelona), Spain} \\
\r 4  {\eightit Istituto Nazionale di Fisica Nucleare, University of Bologna,
I-40127 Bologna, Italy} \\
\r 5  {\eightit Brandeis University, Waltham, Massachusetts 02254} \\
\r 6  {\eightit University of California at Davis, Davis, California  95616} \\
\r 7  {\eightit University of California at Los Angeles, Los 
Angeles, California  90024} \\
\r 8  {\eightit University of California at San Diego, La Jolla, California  92093} \\ 
\r 9  {\eightit University of California at Santa Barbara, Santa Barbara, California 
93106} \\ 
\r {10} {\eightit Instituto de Fisica de Cantabria, CSIC-University of Cantabria, 
39005 Santander, Spain} \\
\r {11} {\eightit Carnegie Mellon University, Pittsburgh, PA  15213} \\
\r {12} {\eightit Enrico Fermi Institute, University of Chicago, Chicago, 
Illinois 60637} \\
\r {13}  {\eightit Joint Institute for Nuclear Research, RU-141980 Dubna, Russia}
\\
\r {14} {\eightit Duke University, Durham, North Carolina  27708} \\
\r {15} {\eightit Fermi National Accelerator Laboratory, Batavia, Illinois 
60510} \\
\r {16} {\eightit University of Florida, Gainesville, Florida  32611} \\
\r {17} {\eightit Laboratori Nazionali di Frascati, Istituto Nazionale di Fisica
               Nucleare, I-00044 Frascati, Italy} \\
\r {18} {\eightit University of Geneva, CH-1211 Geneva 4, Switzerland} \\
\r {19} {\eightit Glasgow University, Glasgow G12 8QQ, United Kingdom}\\
\r {20} {\eightit Harvard University, Cambridge, Massachusetts 02138} \\
\r {21} {\eightit The Helsinki Group: Helsinki Institute of Physics; and Division of
High Energy Physics, Department of Physical Sciences, University of Helsinki, FIN-00044, Helsinki, Finland}\\
\r {22} {\eightit Hiroshima University, Higashi-Hiroshima 724, Japan} \\
\r {23} {\eightit University of Illinois, Urbana, Illinois 61801} \\
\r {24} {\eightit The Johns Hopkins University, Baltimore, Maryland 21218} \\
\r {25} {\eightit Institut f\"{u}r Experimentelle Kernphysik, 
Universit\"{a}t Karlsruhe, 76128 Karlsruhe, Germany} \\
\r {26} {\eightit High Energy Accelerator Research Organization (KEK), Tsukuba, 
Ibaraki 305, Japan} \\
\r {27} {\eightit Center for High Energy Physics: Kyungpook National
University, Taegu 702-701; Seoul National University, Seoul 151-742; and
SungKyunKwan University, Suwon 440-746; Korea} \\
\r {28} {\eightit Ernest Orlando Lawrence Berkeley National Laboratory, 
Berkeley, California 94720} \\
\r {29} {\eightit University of Liverpool, Liverpool L69 7ZE, United Kingdom} \\
\r {30} {\eightit University College London, London WC1E 6BT, United Kingdom} \\
\r {31} {\eightit Massachusetts Institute of Technology, Cambridge,
Massachusetts  02139} \\   
\r {32} {\eightit Institute of Particle Physics, McGill University,
Montr\'{e}al, Canada H3A~2T8; and University of Toronto, Toronto, Canada
M5S~1A7} \\
\r {33} {\eightit University of Michigan, Ann Arbor, Michigan 48109} \\
\r {34} {\eightit Michigan State University, East Lansing, Michigan  48824} \\
\r {35} {\eightit Institution for Theoretical and Experimental Physics, ITEP,
Moscow 117259, Russia} \\
\r {36} {\eightit University of New Mexico, Albuquerque, New Mexico 87131} \\
\r {37} {\eightit Northwestern University, Evanston, Illinois  60208} \\
\r {38} {\eightit The Ohio State University, Columbus, Ohio  43210} \\  
\r {39} {\eightit Okayama University, Okayama 700-8530, Japan}\\  
\r {40} {\eightit Osaka City University, Osaka 588, Japan} \\
\r {41} {\eightit University of Oxford, Oxford OX1 3RH, United Kingdom} \\
\r {42} {\eightit University of Padova, Istituto Nazionale di Fisica 
          Nucleare, Sezione di Padova-Trento, I-35131 Padova, Italy} \\
\r {43} {\eightit University of Pennsylvania, Philadelphia, 
        Pennsylvania 19104} \\   
\r {44} {\eightit Istituto Nazionale di Fisica Nucleare, University and Scuola
               Normale Superiore of Pisa, I-56100 Pisa, Italy} \\
\r {45} {\eightit University of Pittsburgh, Pittsburgh, Pennsylvania 15260} \\
\r {46} {\eightit Purdue University, West Lafayette, Indiana 47907} \\
\r {47} {\eightit University of Rochester, Rochester, New York 14627} \\
\r {48} {\eightit The Rockefeller University, New York, New York 10021} \\
\r {49} {\eightit Istituto Nazionale di Fisica Nucleare, Sezione di Roma 1,
University di Roma ``La Sapienza," I-00185 Roma, Italy}\\
\r {50} {\eightit Rutgers University, Piscataway, New Jersey 08855} \\
\r {51} {\eightit Texas A\&M University, College Station, Texas 77843} \\
\r {52} {\eightit Texas Tech University, Lubbock, Texas 79409} \\
\r {53} {\eightit Istituto Nazionale di Fisica Nucleare, University of Trieste/\
Udine, Italy} \\
\r {54} {\eightit University of Tsukuba, Tsukuba, Ibaraki 305, Japan} \\
\r {55} {\eightit Tufts University, Medford, Massachusetts 02155} \\
\r {56} {\eightit Waseda University, Tokyo 169, Japan} \\
\r {57} {\eightit Wayne State University, Detroit, Michigan  48201} \\
\r {58} {\eightit University of Wisconsin, Madison, Wisconsin 53706} \\
\r {59} {\eightit Yale University, New Haven, Connecticut 06520} \\
\end{center}
  
%
%
\begin{abstract}
We present the results of a search for doubly-charged Higgs bosons
($H^{\pm\pm}$) decaying to dileptons using $\approx 240$ pb$^{-1}$ of
$p{\bar p}$ collision data collected by the CDF II experiment at the Fermilab
Tevatron.  In our search region, given by
 same-sign dilepton mass $m_{ll'} > 80$ GeV/$c^2$ (100 GeV/$c^2$ for
 dielectron channel),
  we observe no evidence for doubly-charged Higgs production.
 We set limits on $\sigma (p{\bar p}
\rightarrow H^{++}H^{--} \rightarrow l^+l^+l^-l^-)$ as a function of the
mass of the doubly-charged Higgs boson
 and the chirality
of its couplings.  Assuming exclusive same-sign dilepton decays,
 we derive lower mass limits on $H^{\pm \pm}_L$
of        133 GeV/$c^2$, 136 GeV/$c^2$, and 115 GeV/$c^2$
 in the $ee$, $\mu\mu$, and $e\mu$ channels, respectively,
        and a lower mass limit of 113 GeV/$c^2$ on $H^{\pm \pm}_R$
 in the $\mu\mu$ channel, all at the 95\% confidence level.
\end{abstract}

\vfill\eject

The standard model (SM) 
 gives a good description of the known fundamental particles, 
using the $SU(3)_C \times SU(2)_L \times U(1)_Y$ gauge
 group to describe   their non-gravitational 
 interactions. 
 The $SU(2)_L \times U(1)_Y$ electroweak 
 gauge symmetry is broken to $U(1)_{EM}$ 
 by the Higgs mechanism, but
 a Higgs boson has yet to be observed.  The observation of any Higgs 
particle would be an important step toward understanding the physics 
at the electroweak scale.  In addition to the SM 
$SU(2)_{L}$ Higgs doublet, a number of models \cite{smext,lrsym,lighth++} 
predict new Higgs doublets or triplets containing doubly-charged Higgs 
bosons ($H^{\pm\pm}$).  For example, the left-right 
 symmetric model \cite{lrsym}, 
predicated on a right-handed version of the weak force $SU(2)_{R}$, 
requires a Higgs triplet.  The model predicts 
light neutrino masses by the seesaw mechanism \cite{lightnu}, consistent
 with recent data on neutrino oscillations~\cite{oscillations}. 
 Furthermore, the left-right 
 symmetric model suggests light 
 (${\cal O}$(100 GeV/$c^2$)) doubly-charged Higgs
 particles if supersymmetry is 
a property of nature \cite{lighth++}, and is therefore of interest for
 direct searches at high-energy colliders.
\par
Doubly-charged Higgs bosons couple directly to leptons, photons, $W$ and $Z$ 
bosons, and singly-charged Higgs bosons ($H^{\pm}$).  
 The $H_L^{\pm\pm}$ and $H_R^{\pm\pm}$ bosons 
 respectively couple to left- and right-handed particles,
and may have different fermionic couplings.
   Their coupling to a pair of $W$ bosons is experimentally constrained to be small due to 
the small observed value of $|\rho_{EW}-1|$ \cite{rho}, resulting in a negligible cross 
section for the process $p\bar{p}\rightarrow W^{\pm} \rightarrow W^{\mp}H^{\pm\pm}$.  
Therefore, $H^{\pm\pm}$ production  would be  dominated 
by the reaction $p\bar{p}\rightarrow Z/\gamma^* 
\rightarrow H^{++}H^{--}$, whose cross section is independent of the 
 $H^{\pm\pm}$ fermionic couplings.
\par
The $H^{\pm\pm}$ decays predominantly to charged
leptons if $m_{H^{\pm\pm}}<2m_{H^{\pm}}$ and $m_{H^{\pm\pm}}-m_{H^{\pm}}<
m_{W^{\pm}}$~\cite{schiggs}. The leptonic decays conserve the quantum number
 $B-L$, where $B$ is baryon number and $L$ is lepton number. 
  The $H^{\pm \pm}$ couplings $h_{ll'}$ to electrons and 
 muons are experimentally 
constrained by the absence of $H^{\pm\pm}$ production in $e^+e^-$ 
collisions ($h_{ee}<0.07$) \cite{singlesearch}, 
 and 
the non-observation of the decays $\mu\rightarrow 3e$ ($h_{ee}h_{e\mu}<3.2\times 10^{-7}$) 
and $\mu\rightarrow e \gamma$ ($h_{\mu\mu}h_{e\mu}<2\times 10^{-6}$) 
\cite{coupling}.  The experimental constraints on the couplings 
 (quoted here for $m_{H^{\pm\pm}}=100$ GeV/$c^2$) weaken with
increasing doubly-charged Higgs mass. 
 The $h_{\mu\mu}$ coupling is probed by 
 measurements of the anomalous magnetic moment of the muon 
$(g-2)_{\mu}$; the previous 
 limit $h_{\mu\mu} < 0.25$~\cite{coupling} 
 has not been reanalyzed using
  the most recent $(g-2)_{\mu}$ measurement~\cite{recentg2}. 
\par
Direct searches by the 
OPAL and L3 collaborations in $e^+e^-$ collisions \cite{opall3} have excluded 
doubly-charged Higgs bosons below masses of about 100 GeV/$c^2$, assuming
  exclusive 
$H^{\pm\pm}$ decay to a given dilepton channel.  A recent search by 
 the D\O\ collaboration  in the $\mu\mu$ channel~\cite{d0} has 
 excluded $H_L^{\pm\pm}$ below a mass of 118 GeV/$c^2$.
  In this Letter, we describe a search for 
 doubly-charged resonances in 
 the same-sign $ee$, $e\mu$, and $\mu\mu$ channels, using 
 $\approx 240$ pb$^{-1}$~\cite{luminosity} of data 
 collected at $\sqrt{s} = 1.96$ TeV by the CDF II
 experiment at the Fermilab
 Tevatron. We present our results 
using the doubly-charged Higgs production model~\cite{lightnu}, and set the 
world's highest
  mass limits in the electron and
 muon channels.  We probe the range of coupling $10^{-5} < h_{ll'} < 0.5$,
which corresponds to narrow resonances that decay promptly 
  ($c\tau< 10 \; \mu$m, where $\tau$ is the lifetime).
\par
The CDF II detector~\cite{cdf2} consists of three 
 major subsystems:  an inner tracking detector,
  a lead (iron) scintillator sampling 
calorimeter for measuring electromagnetic (hadronic) showers, and outer drift 
chambers for muon identification.  The inner detector includes a
 high-resolution wire chamber 
(the Central Outer Tracker, or COT~\cite{COT}) which, along with the central 
calorimeter and muon
 system,  covers the pseudorapidity 
 interval $| \; \eta \; | < 1$~\cite{coords}. 
\par 
 Our strategy is to search for one of the pair-produced $H^{\pm \pm}$ bosons 
to maximize the sensitivity, and to permit detection of any singly-produced 
 doubly-charged resonance. 
  The event triggers can be classified by the requirements of (1)
  two energy clusters with $E_T>18$ GeV in the electromagnetic calorimeter (2EM),
 (2) a central electromagnetic cluster with  $E_T>18$ 
 GeV and  matching track $p_{T}>9$ GeV/$c$ (1EM), or (3) a COT track with  
 $p_{T}>18$ GeV/$c$ 
   with an associated track segment (``stub'') 
 in the muon detectors. 
\par
The same-sign $ee$ sample is selected primarily using the 2EM trigger. 
  In the offline
 analysis, we require two same-sign central 
electrons with calorimeter $E_T > 30$ GeV and COT track $p_T > 10$ GeV/$c$. 
 Electrons are identified  
using the ratio of calorimeter energy $(E)$ to track momentum $(p)$ 
 $(\frac {E}{pc} < 4)$, longitudinal and 
lateral shower profiles, track-cluster matching, 
calorimeter isolation energy in a surrounding cone, and photon-conversion
 identification using  
the tracker.  The same-sign $ee$ sample corresponds to an integrated 
luminosity of $(235 \pm 13)$ pb$^{-1}$.  The 
 luminosity is determined by 
measuring the rate of inelastic collisions, and the uncertainty has
 equal contributions from
the uncertainty on the inelastic cross section and the uncertainty on the 
 acceptance of the 
luminosity counters.
\par
The same-sign $\mu\mu$ sample is selected using the single-muon trigger, with
 a consistent offline requirement of a matching stub. 
  We select tracks with $p_T > 25$ GeV/$c$ that are minimum-ionizing, \it i.e. \rm 
 have small electromagnetic and hadronic energy depositions in the
 calorimeters.  The cosmic-ray
  muon background is suppressed by requiring the muons to originate from the 
 beam line, to be 
coincident in time with each other and with a $p\bar{p}$ collision, 
 and to be consistent with a 
pair of outgoing particles \cite{COTcosmic}.  Track-quality requirements 
and calorimeter isolation suppress hadronic-jet
 backgrounds.
   The integrated luminosity of the same-sign $\mu\mu$ sample is $(242 \pm 14)$ 
pb$^{-1}$.
\par
The same-sign $e\mu$ sample is selected mainly using the 1EM trigger.
   We require a central 
electron and a track matched to a muon 
 stub.  The stub requirement significantly reduces 
background, but also reduces the 
 fiducial acceptance of $H^{\pm\pm} \rightarrow e \mu$ relative to the
 $\mu\mu$ and $ee$ samples.  The integrated luminosity of the same-sign $e\mu$
 sample is 
$(240 \pm 14)$ pb$^{-1}$. All electron and muon
  tracks are constrained to the transverse position of the beam to improve
 their momentum
 resolution. 
\par
 We calculate trigger efficiencies using separate 
 unbiased triggers,  the tracking efficiency  using $Z \rightarrow ee$ 
events, and 
  the lepton-identification efficiencies  with 
 $Z \rightarrow ee / \mu\mu$ events.
 We obtain $(96.6 \pm 0.4)$\% and 
  $(100.00^{+0.00}_{-0.02})$\%  as the efficiencies of the 1EM and 2EM
 triggers, respectively. The muon trigger efficiencies, including
  the offline matching-stub
 requirement, are 
 $(77.1 \pm 1.3)$\% and $(93.9 \pm 0.8)$\% for $| \; \eta \;
 |<0.6$ and $0.6<| \; \eta \; |<1$, respectively, each corresponding
 to a separate detector subsystem. 
 The tracking algorithm is highly efficient
 ($>99$\%) for isolated 
 charged particles within the COT fiducial volume.  The lepton-identification 
efficiencies are (92.7$\pm$0.3)\% and (90.8$\pm$0.2)\% for electrons and 
 muons, respectively.  The 
 corresponding efficiencies measured in simulated~\cite{simulation}
  $Z$ events are 
 (89.3$\pm$0.1)\% and (91.3$\pm$0.1)\%.
 The simulated $H^{\pm\pm}$ detection efficiency is corrected by the
 ratio of data to simulated  $Z$ boson efficiencies. 
\par
The potential backgrounds from 
 SM processes are (1)
hadrons that decay to leptons or are misidentified as such, (2) leptonic 
decays 
of $W$ bosons, produced in association with 
 hadronic jet(s) ($W+$jet), (3) $Z/\gamma^*$ 
decays (Drell-Yan), 
where the same-sign track comes from a photon conversion,
 (4) $WZ$ production, where both the $W$ and $Z$ decay leptonically, 
and   (5) cosmic rays.
\par
The hadronic background is estimated using lepton-triggered events with 
two same-sign lepton candidates~\cite{lepton},
 each failing the identification requirements (``failing
 lepton candidate'').  The ratio
 of the number of lepton candidates passing to the number
  failing the requirements (the ``pass-fail ratio'') is measured using 
 jet data samples.     These samples are selected either using high-$ E_T$ 
($>100$ GeV) or low-$E_T$ 
($>20$ GeV) jet triggers, or using single-lepton triggers and excluding 
leptonic $W$ and $Z$ decays.  The pass-fail ratio is ${\cal O}(0.05)$, with a
 systematic uncertainty of $\approx 80$\% arising from its
 sample dependence.  It is used to apply a weight to
 each candidate lepton (as a function of $E_T$) in 
events with two failing lepton candidates
 to obtain the dilepton mass distribution.
\par
The $W+$jet background is determined by applying the pass-fail ratio as a 
 weight to $W$ data events which 
 have a second 
failing lepton and $25 < \met < 60$~GeV.  The expected misidentified-$W$
  contribution (from jets) is
   subtracted to prevent 
 double-counting. We use simulated~\cite{simulation} $W$+jet events
 to correct for the acceptance of the \met\ requirement.  Background from
 $W\gamma$
 production, where the photon converts to an $e^+e^-$ pair, is implicitly
 included in this estimate. It is 
 studied explicitly using the simulation and
 found to be negligible. 

\begin{figure}[!htbp]
\begin{center}
\begin{minipage}[ht]{7.0cm}
\epsfysize = 7.0cm
\epsffile{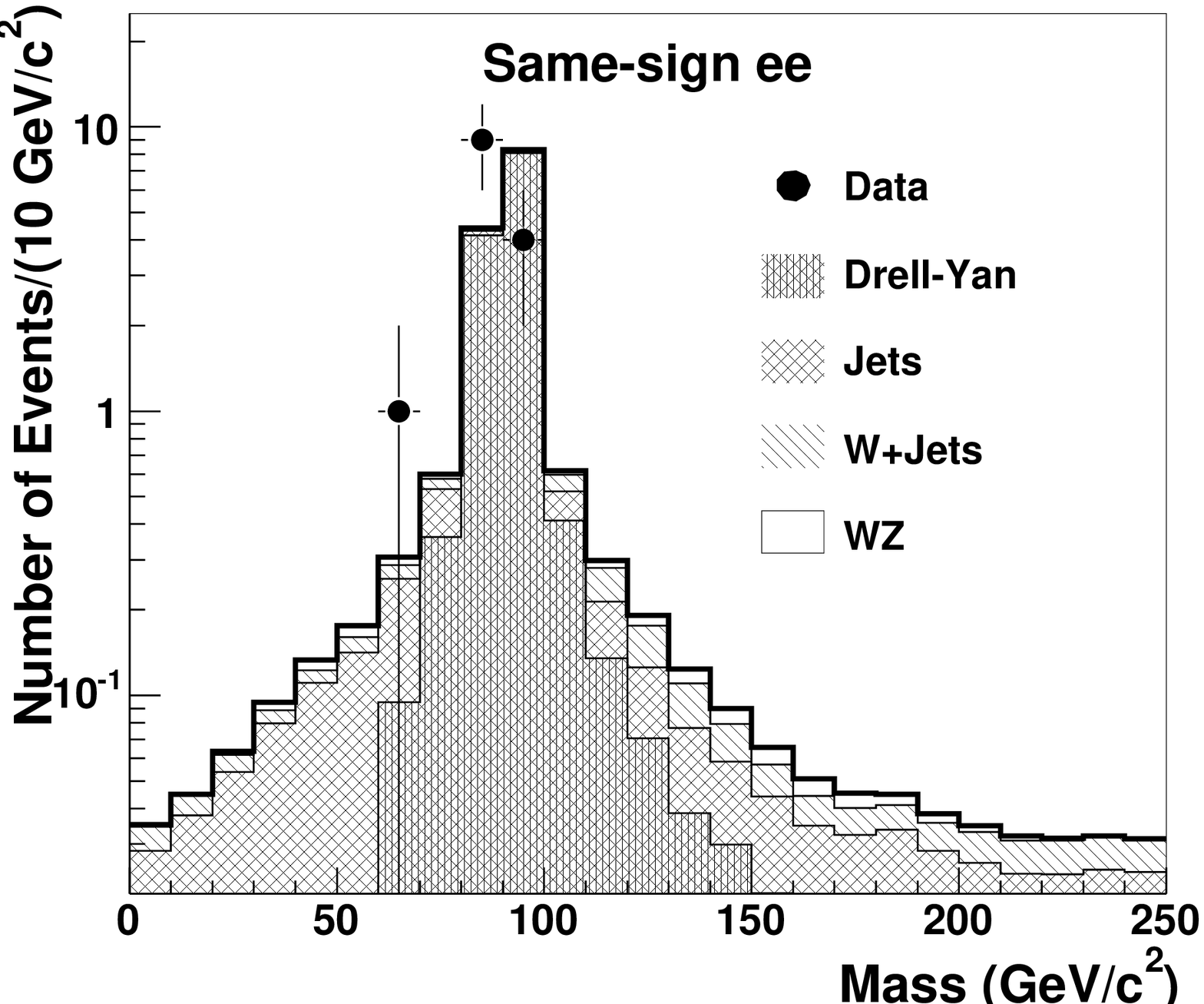}
\end{minipage}
\hskip 0.3in
\begin{minipage}[ht]{7.0cm}
\epsfysize = 7.0cm
\epsffile{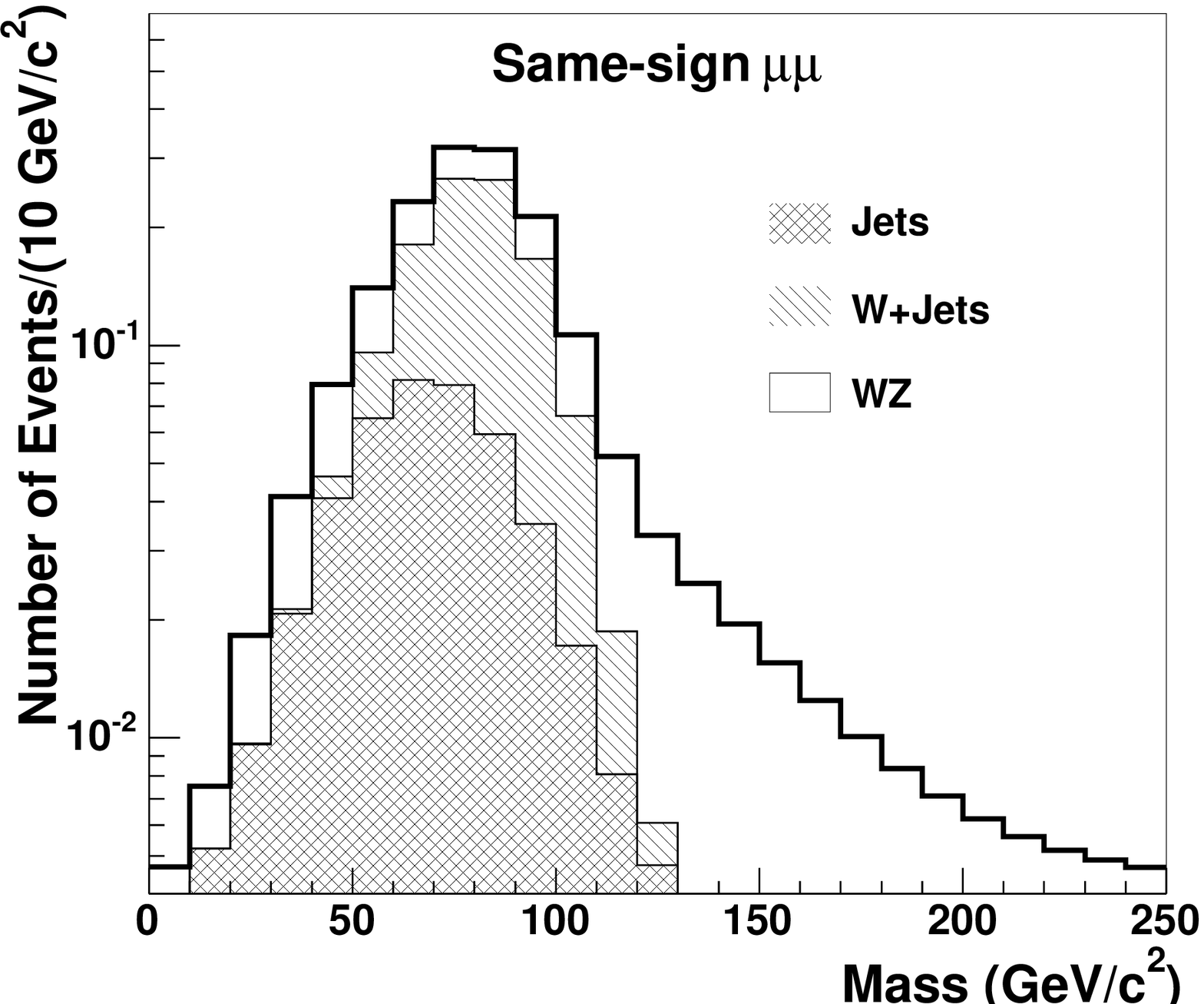}
\end{minipage}
\begin{minipage}[ht]{7.0cm}
\epsfysize = 7.0cm
\epsffile{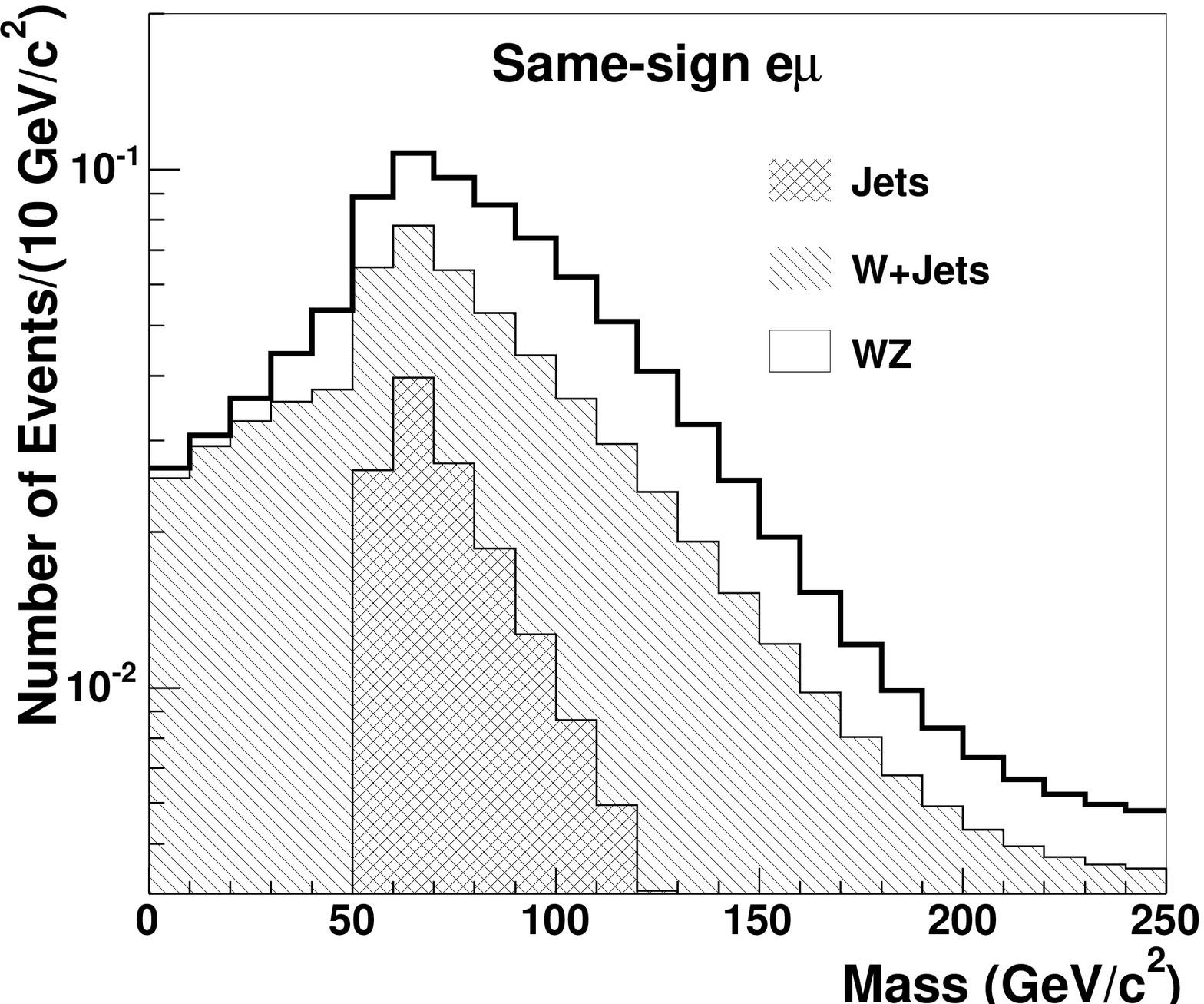}
\end{minipage}
\vspace*{3mm}
\caption{The same-sign 
 dilepton mass distributions of the $ee$ data and the 
 cumulative SM contributions
 to the $ee$ (top-left), $\mu\mu$ (top-right), and $e\mu$ (bottom) samples.
   The 
solid line is the overall sum of the indicated areas. No same-sign 
$\mu\mu$ or $e\mu$ events are observed.
 }
\label{fig:totbd}
\end{center}
\end{figure}

\par
Background from $Z/\gamma^*\rightarrow e^+e^-$ occurs when one 
 electron radiates a 
photon which subsequently converts to an $e^+e^-$ pair.  When a same-sign 
 conversion 
electron has higher momentum than the prompt electron and is associated 
 with the cluster, 
 the event is reconstructed with two same-sign electrons.  The 
mass dependence  is obtained from simulated~\cite{simulation}
 Drell-Yan events. The simulated sample is
 normalized using the number of same-sign  candidates in the $Z$
 mass region (80 GeV/$c^2 < m_{ee} < 100$ GeV/$c^2$), after subtracting jet
 and $W$+jet 
contributions.  
\par
Background from $WZ \rightarrow l\nu ll$ 
 production is estimated using simulation~\cite{simulation}.  We use the 
 next-to-next-to-leading order 
production cross section of 4.0 pb \cite{wzxsec}, and apply the trigger, 
 tracking, 
and lepton-identification efficiencies to the events that pass the kinematic 
 and 
geometric selection. 
\par
The cosmic-ray background is estimated using COT timing information.  
 We use an
     independently identified sample of cosmic rays to estimate the residual
     contribution surviving the timing requirements made in the
     $\mu\mu$ analysis. 
 The expected cosmic-ray background is found to be $0.02 \pm 0.02$
events, which we take to be negligible.
\par

\begin{table}[ht]
\begin{center}
\begin{tabular}{ccc}
Background & Low-Mass Region & High-Mass Region  \\
\hline  $Z / \gamma^* \rightarrow ee$ &  $0.46 \pm 0.13$
				&  $0.37 \pm 0.11$ \\
  	Jets$\rightarrow ee$   	&  $0.47^{+0.23}_{-0.19} $
				&  $0.62 ^{+0.71}_{-0.44} $ \\
	$W+$jet$\rightarrow ee$   	&  $0.14 \pm 0.08$
				&  $0.36 \pm 0.21$ \\
  	$ WZ \rightarrow ee$   	&  $0.07 \pm 0.02$
				&  $0.11 \pm 0.03$ \\
\hline  Total $ee$   	&  $1.1 \pm 0.4$ &  $1.5^{+0.9}_{-0.6} $ \\
\hline
\hline  Jets$\rightarrow\mu\mu$ & $0.30 ^{+0.24}_{-0.16} $
		    &  $0.19^{+0.35}_{-0.17} $ \\
  	$W+$jet$\rightarrow\mu\mu$  & $0.32 \pm 0.22$
		    &  $0.40 \pm 0.27$ \\
 	$WZ \rightarrow\mu\mu$ & $0.21 \pm 0.04$
		    &  $0.19 \pm 0.03$ \\
\hline  Total $\mu\mu$ & $0.8 \pm 0.4$
		    &  $0.8^{+0.5}_{-0.4} $ \\
\hline
\hline  Jets$\rightarrow e\mu$ & $0.09 \pm 0.05$
		    &   $0.06 \pm 0.05$ \\
  	$W+$jet$\rightarrow e \mu$ & $0.22^{+0.24}_{-0.15} $
		    &  $0.25 \pm 0.17$ \\
 	$WZ \rightarrow e \mu$ & $0.12 \pm 0.02$
		    &  $0.12 \pm 0.03$ \\
\hline  Total $e \mu$ & $0.4 \pm 0.2$
		    &  $0.4 \pm 0.2$ \\
\end{tabular}
\end{center}
\vskip 0.1in
\caption{The integrated background for the $ee$, $\mu\mu$ and $e \mu$
samples for the 
low-mass ($< 80$ GeV/$c^2$) and high-mass (100-300 GeV/$c^2$ 
 for $ee$, 80-300 GeV/$c^2$ for
$\mu\mu$ and $e \mu$)
regions. } 
\label{tbl:totalregion}
\end{table}

Figure~\ref{fig:totbd} shows the total background and the data as a 
function of $m_{ll'}$ for each sample.  The predominantly 
 back-to-back lepton topologies, 
 the kinematic thresholds, and the typical lepton $p_T$ from
$W$  or $Z$ decays lead to the observed peaked shapes of the
 background distributions.
 The search is performed 
 in the region of $m_{ll'}>80$ GeV/$c^2$ for the $\mu\mu$ and  $e\mu$ 
samples, and in the region of $m_{ee}>100$ GeV/$c^2$ for the $ee$ sample.  The
  low-mass 
regions ($m_{ll'}<80$ GeV/$c^2$) are used to check our background predictions.
 Table~\ref{tbl:totalregion} summarizes the total
 background predictions.    We 
estimate 1.1$\pm$0.4 ($ee$), 0.8$\pm$0.4 ($\mu\mu$), and 0.4$\pm$0.2 ($e\mu$) 
events in the low-mass regions, and observe one $ee$ event
  ($m_{ee} = 70$ GeV/$c^2$) 
and no $\mu\mu$ or $e\mu$ events.  As an additional check, we compare the 
 predicted 
and observed backgrounds for same-sign dilepton events with 
 one failing lepton candidate and $\met <15$ GeV.  The expectations of 
$54 \pm 21$ ($ee$), $7.6 \pm 3.1$ ($\mu\mu$), and $2.4 \pm 0.8$
 ($e \mu$) events are  
consistent with the observed numbers of 63 ($ee$), 8 ($\mu\mu$), and 
2 ($e \mu$) events.
\par

\begin{figure}[!htbp]
\begin{center}
\epsfysize = 7.2cm
\hspace*{0.1cm}
\epsffile{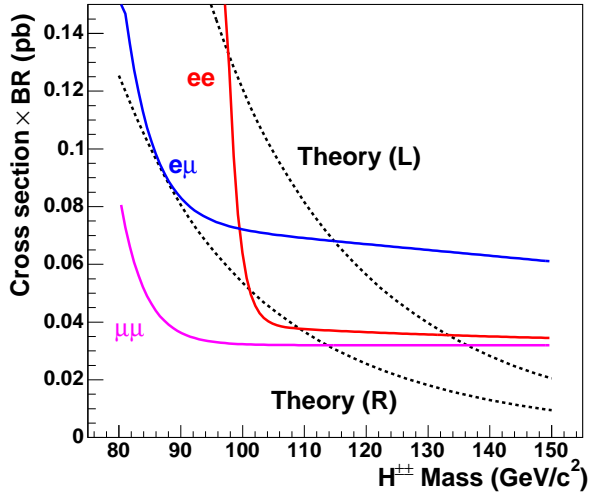}
\vspace*{3mm}
\caption{Experimental limits on cross section~$\times$~branching ratio
 at 95\% C.L. as a function of 
doubly-charged Higgs mass (solid curves).  Dotted curves show the 
 theoretical  next-to-leading order total cross sections~\protect\cite{xsec} 
for left-handed and right-handed $H^{\pm\pm}$ couplings. }
\label{fig:limit}
\end{center}
\end{figure}

The same-sign dilepton mass resolution is $\approx 3.5$\% of the mass.  The 
 intrinsic 
$H^{\pm\pm}$ width is equal to $h^2_{ll'}m_{H^{\pm\pm}}/8\pi$ \cite{rho}, and 
contributes negligibly to the reconstructed mass.  We define  search 
windows of $\pm$10\% of a given $H^{\pm\pm}$ mass, corresponding to a
  $\pm 3\sigma$ window. 
  We predict the 
  acceptances as a function of $H^{\pm \pm}$ mass
 using the simulation~\cite{simulation},
 including the efficiency scale factors.  The acceptance 
 systematic
uncertainty is dominated by the parton distribution function uncertainty, 
 which we estimate
to be 4\% using the MRST prescription~\cite{mrst}. 
 In the mass range of interest, the acceptances are 
  $\approx 34$\% for the 
 $ee$ and $\mu \mu$ channels
 and  $\approx 18$\% for the $e \mu$ channel. 
\par
\begin{figure}[!htbp]
\begin{center}
\epsfysize = 7.2cm
\hspace*{0.1cm}
\epsffile{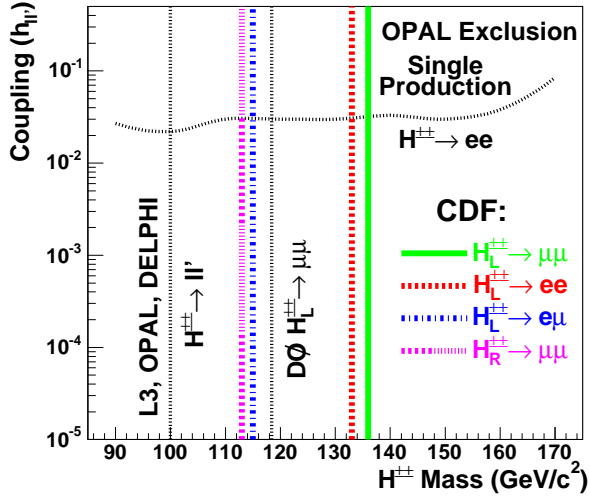}
\vspace*{3mm}
\caption{The doubly-charged Higgs lower mass limits versus lepton coupling 
($h_{ll'}$) from this analysis, assuming 
exclusive decay to a given dilepton pair. Our limits are valid for
  $h_{ll'} > 10^{-5}$.
  Previous limits~\protect\cite{singlesearch,opall3,d0} are also shown. }
\label{fig:massvh}
\end{center}
\end{figure}

No events are found in the high-mass regions of the $ee$, $\mu\mu$ and $e\mu$ 
samples.  This null result yields a 95\% confidence
 level (C.L.) upper limit on the cross section 
 as a function of doubly-charged Higgs mass (Fig.~\ref{fig:limit}).
  We calculate
the limit using a Bayesian method~\cite{bayes} with a flat prior for the 
signal and Gaussian priors for background and acceptance uncertainties.  
 Through 
comparison with the theoretical cross sections~\cite{xsec}, we obtain mass
  limits 
 of 133 GeV/$c^2$, 136 GeV/$c^2$, and 115 GeV/$c^2$, for exclusive
  $H_L^{\pm\pm}$ decays 
to $ee$, $\mu\mu$, and $e\mu$, respectively, and 113 GeV/$c^2$ for exclusive 
$H_R^{\pm\pm}$ decays to $\mu\mu$. 
  Figure~\ref{fig:massvh} shows these results
in the mass-coupling plane, along with the
 current world limits.

In summary, we have performed an inclusive search for doubly-charged 
 resonances in 
same-sign $ee$ data with $m_{ee} > 100$ GeV/$c^2$, and same-sign
 $\mu\mu$ and $e\mu$ data with $m_{ll'} > 80$ GeV/$c^2$.  We have found no 
 evidence for 
new doubly-charged resonances, and have significantly extended the existing 
 mass limits 
on doubly-charged Higgs bosons decaying exclusively to $ee$ ($m_{H_L^{\pm\pm}} > 133$ 
GeV/$c^2$), $\mu\mu$ ($m_{H_L^{\pm\pm}} > 136$ GeV/$c^2$ and $m_{H_R^{\pm\pm}} > 113$ GeV/$c^2$), 
or $e\mu$ ($m_{H_L^{\pm\pm}} > 115$ GeV/$c^2$) final states.
 
We thank M. M$\rm\ddot{u}$hlleitner and M. Spira for calculating the next-to-leading
 order  $H^{\pm \pm}$ production cross section.
We thank the Fermilab staff and the technical staffs of the participating institutions for their vital contributions.  This work was supported by the U.S. Department of Energy and National Science Foundation; the Italian Istituto Nazionale di Fisica Nucleare; the Ministry of Education, Culture, Sports, Science and Technology of Japan; the Natural Sciences and Engineering Research Council of Canada; the National Science Council of the Republic of China; the Swiss National Science Foundation; the A.P. Sloan Foundation; the Bundesministerium f\"ur Bildung und Forschung, Germany; the Korean Science and Engineering Foundation and the Korean Research Foundation; the Particle Physics and Astronomy Research Council and the Royal Society, UK; the Russian Foundation for Basic Research; the Comision Interministerial de Ciencia y Tecnologia, Spain; work supported in part by the European Community's Human Potential Programme under contract HPRN-CT-20002, Probe for New Physics; and this work was supported by Research Fund of Istanbul University Project No. 1755/21122001.


\begin{references}
\bibitem{coords}
CDF uses a cylindrical coordinate system in which $\phi$ is the azimuthal angle, $r$ is 
the radius from the nominal beamline, and $+z$ points in the direction of the proton
beam and is zero at the center of the detector.  The pseudorapidity
 $\eta = -$ln[$\tan(\theta/2)$], where $\theta$ is the polar angle with respect
 to the $z$ axis.  Calorimeter energy (track momentum) measured transverse to
  the beam is denoted 
as $E_T$ ($p_T$), and the total calorimetric transverse energy 
imbalance is denoted as \met.
\bibitem{lepton}
Electron candidates are EM clusters with $E_T > 25$ GeV and matching track
 $p_T > 10$ GeV/$c$, and muon
 candidates are tracks with $p_T > 20$ GeV/$c$.
\bibitem{simulation}
All simulated samples are generated with {\sc pythia}~\cite{pythia} and use a
 detector simulation based on  {\sc geant}~\cite{geant}.
\end{references}
\end{document}